\documentclass[twocolumn,reprint,amsmath,amssymb,aps,showpacs,superscriptaddress,prb]{revtex4-1}

\usepackage{graphicx}
\usepackage{bm}
\usepackage{epsfig}
\usepackage{amssymb}
\usepackage{amsfonts}
\usepackage{braket}
\usepackage{color}
\usepackage{epstopdf}
\usepackage{hyperref}
\usepackage{float}
\restylefloat{table}
\usepackage{bibentry}
\usepackage{color}
\usepackage{multirow}
\usepackage[caption=false]{subfig}
\usepackage{ulem}

\newcommand{\bpm}{\begin{pmatrix}}
\newcommand{\epm}{\end{pmatrix}}
\newcommand{\ba}{\begin{eqnarray}}
\newcommand{\ea}{\end{eqnarray}}
\newcommand{\bd}{\begin{displaymath}}

\graphicspath{{figures/}}

\begin{document}
\title{Hump-like structure in Hall signal from SrRuO$_3$ ultra-thin films without inhomogeneous anomalous Hall effect}

\author{Byungmin Sohn}
\affiliation{Department of Physics and Astronomy, Seoul National University, Seoul 08826, Korea}
\affiliation{Center for Correlated Electron Systems, Institute for Basic Science, Seoul 08826, Korea}
\author{Bongju Kim}
\affiliation{Department of Physics and Astronomy, Seoul National University, Seoul 08826, Korea}
\affiliation{Center for Correlated Electron Systems, Institute for Basic Science, Seoul 08826, Korea}
\author{Jun Woo Choi}
\affiliation{Center for Spintronics, Korea Institute of Science and Technology, Seoul 02792, Korea}
\author{Seo Hyoung Chang}
\affiliation{Department of Physics, Chung-Ang University, Seoul 06974, Korea}
\author{Jung Hoon Han}
\affiliation{Department of Physics, Sungkyunkwan University, Suwon 16419, Korea}
\author{Changyoung Kim}
\email[Electronic address:$~~$]{changyoung@snu.ac.kr}
\affiliation{Department of Physics and Astronomy, Seoul National University, Seoul 08826, Korea}
\affiliation{Center for Correlated Electron Systems, Institute for Basic Science, Seoul 08826, Korea}\date{\today}

\begin{abstract}
A controversy arose over the interpretation of the recently observed hump features in Hall resistivity $\rho_{xy}$ from ultra-thin SrRuO$_3$ (SRO) film; it was initially interpreted to be due to topological Hall effect but was later proposed to be from existence of regions with different anomalous Hall effect (AHE). In order to settle down the issue, we performed Hall effect as well as magneto-optic Kerr-effect measurements on 4 unit cell SRO films grown on SrTiO$_3$ (001) substrates. Clear hump features are observed in the measured $\rho_{xy}$, whereas neither hump feature nor double hysteresis loop is seen in the Kerr rotation which should be proportional to the magnetization. In addition, magnetization measurement by superconducting quantum interference device shows no sign of multiple coercive fields. These results show that inhomogeneous AHE alone cannot explain the observed hump behavior in $\rho_{xy}$ data from our SRO ultra-thin films. We found that emergence of the hump structure in $\rho_{xy}$ is closely related to the growth condition, high quality films having clear sign of humps.
\end{abstract}
\maketitle

\section*{1. Introduction}
Distinct hump-like features in Hall resistivity near the coercive field are often attributed to the so-called topological Hall effect (THE) (Fig. 1a). Formation of chiral or non-coplanar arrangement of spins can provide fictitious magnetic field or real-space Berry curvature, leading to the THE~\cite{Mn3Sn_1,Mn3Sn_2}. A notable example for such chiral spin structure is the magnetic skyrmions. In fact, THE has been observed in many bulk materials that are known to have magnetic skyrmions~\cite{neubauer09,kanazawa11,yokouchi14,yufan13}. Skyrmion phases have been also obtained in magnetic thin films with heavy metal capping layer, utilizing the naturally occurring broken inversion symmetry as well as large spin-orbit coupling (SOC)~\cite{panagopoulos,granville17}. Later on, a similar approach has been made on oxide thin films and THE-like behavior (hump structures in the Hall data) was indeed observed in systems such as EuO thin film and SrRuO$_3$/SrIrO$_3$ heterostructure~\cite{matsuno,ohuchi}. Then, it was found that SrRuO$_3$ (SRO) ultra-thin films without SrIrO$_3$ capping layer also show THE-like features~\cite{sohn18,kan18,lingfei18,tsinghua18}. However, different causes were given for the THE-like feature in the reports and, as a result, the origin of the THE-like feature remains controversial.

Overall, aside from detailed differences, there are two categories of proposals on the origin of the THE-like feature observed in SRO thin films: one with topological origin (Fig. 1b) and the other from inhomogeneity in anomalous Hall effect (AHE) (Fig. 1c). In the first view, THE from the formation of chiral spin phase is responsible for the hump structures, which is a natural extension of earlier studies on other materials/systems~\cite{panagopoulos,granville17,matsuno,ohuchi}. In the other view, it is argued that the THE-like behavior can also be explained by the sum of different AHEs~\cite{kan18,caviglia18,rutgers18}, suggesting the THE-like feature is not from THE. One of the possible scenarios within the view, as suggested in Ref. \onlinecite{rutgers18}, is that the intrinsic thickness inhomogeneity which inevitably appears due to the step-flow growth mode of SRO thin film~\cite{step-flow} leads to regions with different magnetic properties (thus different AHEs). Another possibility proposed in Ref. \onlinecite{caviglia18} is that the inhomogeneous AHE comes from existence of interfaces with different magnetic properties.

Whether the hump structure comes from THE or inhomogeneity in AHE is an important and controversial issue. However, as both of the two mechanisms can account for the hump features in Hall data, Hall data alone cannot settle down the issue and one therefore needs to rely on other experimental data. In order to resolve the issue, we turn to magnetic measurement of 4 unit cell (u.c.) SRO thin films grown on SrTiO$_3$ (STO) (001) substrate which possess distinct hump features in the Hall data. Magnetic characterization results show that the second view, $i.e.$ superposition of different AHEs alone cannot explain the hump feature in the Hall effect data. Furthermore, we found that the hump behavior is highly sensitive to the residual resistivity ratio (RRR) of the films which is strongly correlated with the growth condition such as the laser fluence for the growth. We suggest that the inconsistent results from various groups might be due to the sensitivity of Ru contents in SRO thin film to the growth condition.

\begin{figure*}[htbp]
\includegraphics[width=1\textwidth]{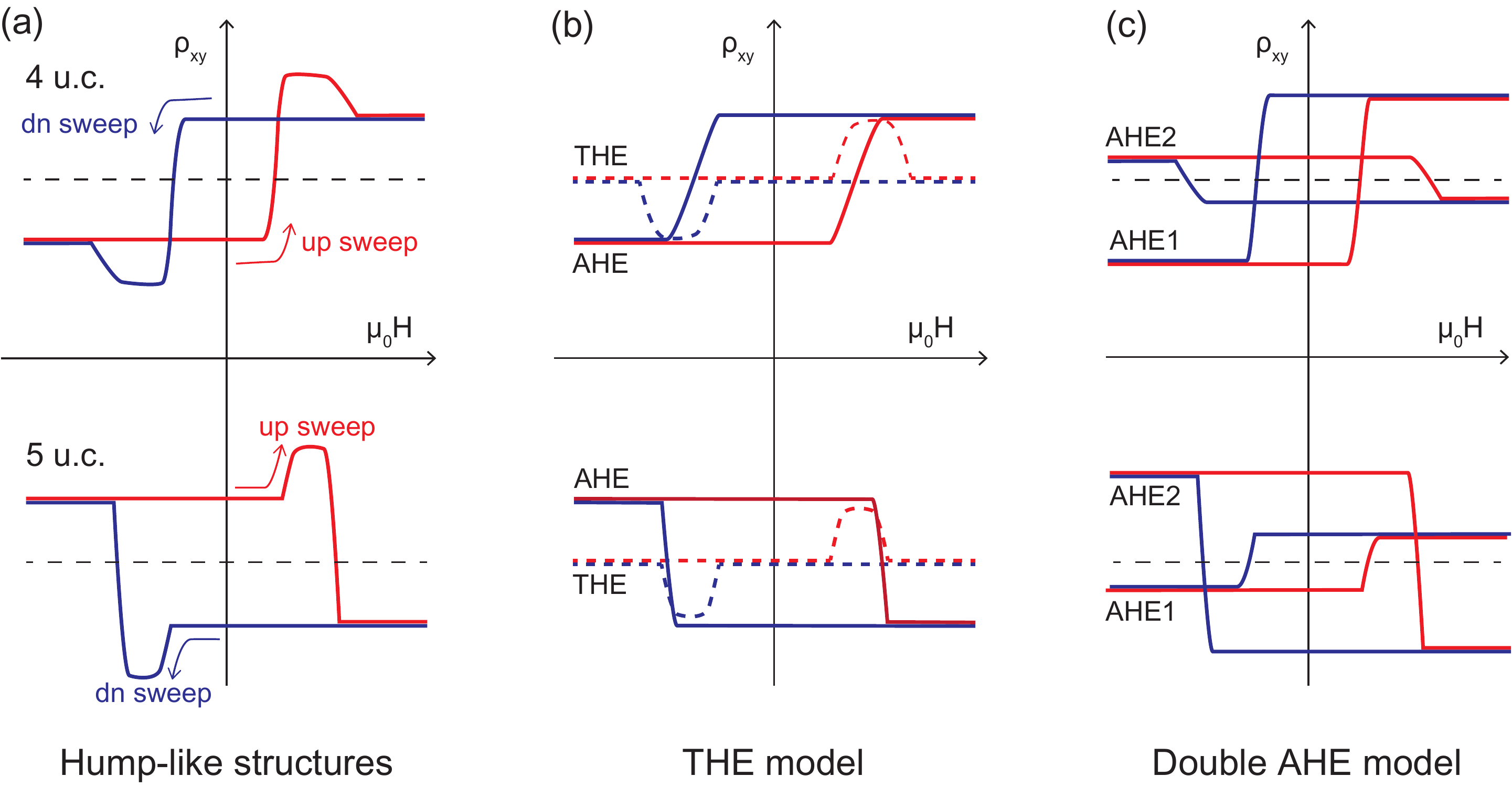}
\caption{Schematics of Hall measurement results for (a) 4 and 5 unit-cell (u.c.) SrRuO$_3$ (SRO) films on SrTiO$_3$ (STO) substrate. The red (blue) line indicates the positive (negative) sweep direction of the magnetic field. Distinct hump-like features exist near the coercive fields. (b) Interpretation of the hump-like feature the sum of topological Hall effect (THE) from non-coplanar spin arrangement (solid) and anomalous Hall effect (AHE) (dotted) for 4 and 5 unit-cell films, respectively. The sign of AHE is different for the two systems. (c) Interpretation the same results in terms of inhomogeneous AHEs. The superposition of two different AHEs with different coercive fields can result in hump-like behaviors.}
\label{fig:1}
\end{figure*}

\section*{2. Methods}
SRO ultra-thin films were grown on TiO$_2$-terminated STO single crystal substrates by using pulsed laser deposition (PLD) technique. TiO$_2$-terminated STO single crystal substrates were prepared by deionized water etching and $in$-$situ$ pre-annealing at $1070\,^{\circ}{\rm C}$ for 30 minutes with an oxygen partial pressure (PO$_2$) of $5\times$10$^{-6}$ Torr. We deposited epitaxial SRO thin film in an oxygen partial pressure of PO$_2$=100 mTorr and the growth temperature of the STO substrate was $700\,^{\circ}{\rm C}$. A KrF Excimer laser (wavelength = 248 nm) was delivered on stoichiometric SRO target with a fluence of 1-2 J/cm$^{2}$ and repetition rate of 2 Hz. Reflection high-energy electron diffraction (RHEED) was used to monitor the growth dynamics.

For the Hall effect measurement of SRO thin films, we prepared $60$ nm-thick Au electrode on top of SRO thin films with a Hall bar geometry by using an electron beam evaporator. Electric transport measurement was carried out by using a Physical Property Measurement System (PPMS), Quantum Design Inc.. The magnetic characterization was performed with a superconducting quantum interface device (SQUID) magnetometry with our-of-plane geometry. We also carried out polar magneto-optic Kerr effect (MOKE) measurement with a laser wavelength of 408 nm in order to measure the out-of-plane magnetization. The laser spot size was $\sim 3$ $\mu$m while the probing depth was $\sim 10$ nm which is large enough that the magnetic signal is from all the SRO layers. Since the magnetic Kerr rotation is expected to be proportional to the magnetization of the thin film~\cite{bader91,kerr,matsuno,xia09}, the magnetic property of the SRO thin film can be obtained from the MOKE result. Because the easy axis of SRO ultrathin film is perpendicular to the thin film on STO (001) substrate, we measured out-of-plane magnetization by SQUID and MOKE measurements ~\cite{klein96, Moty-easy}.

\begin{figure}[htbp]
\includegraphics[width=0.48\textwidth]{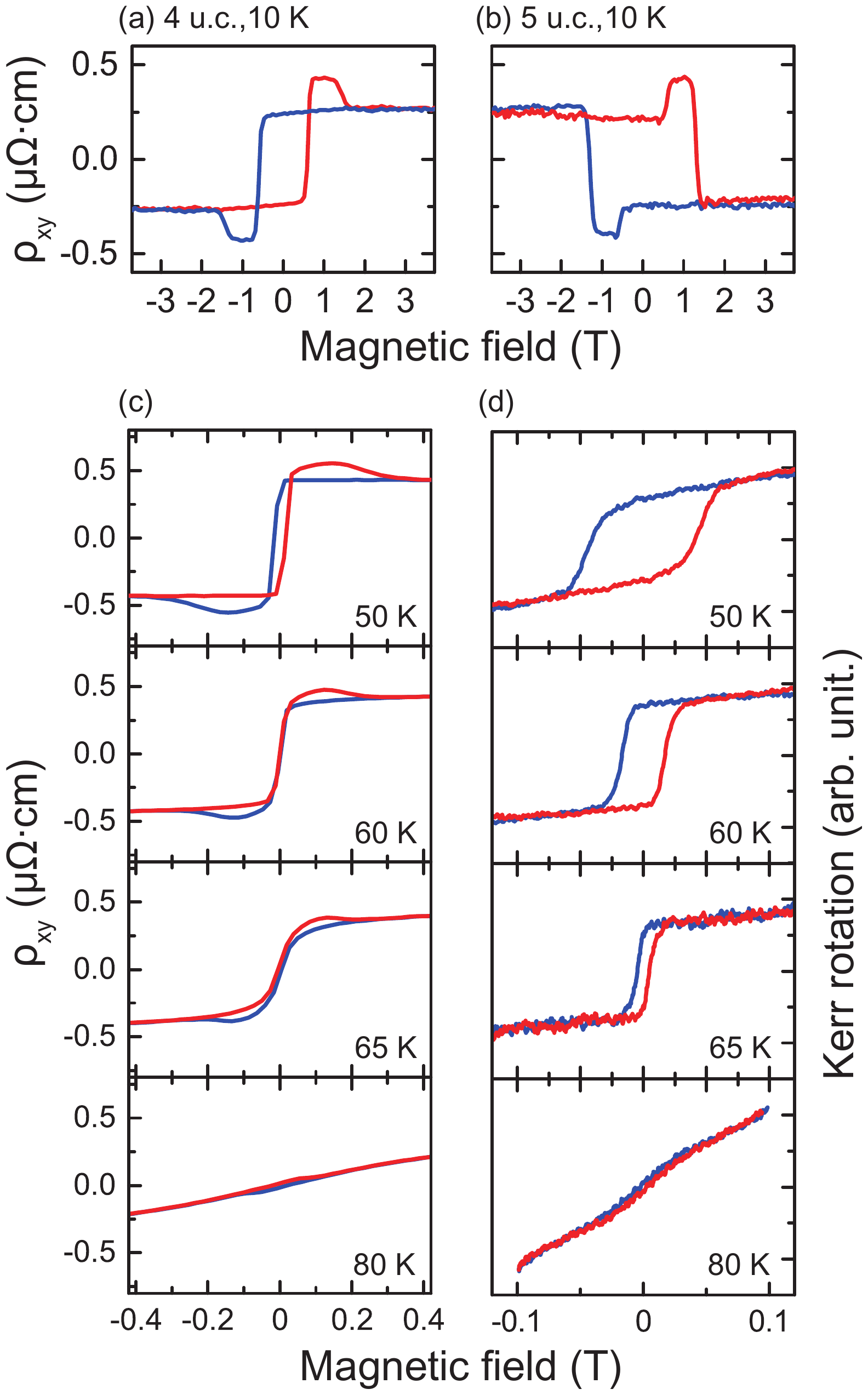}
\caption{Hall measurement results on (a) 4 and (b) 5 u.c. SRO films as a function of the applied magnetic field at 10 K. The linear ordinary Hall effect has been subtracted from the data. The AHE sign is opposite between 4 and 5 u.c. SRO films and clear hump structures are observed for both films. (c) Hall effect measurements of 4 u.c. SRO film at 50, 60, 65, and 80 K. Humps are observed at 50, 60, and 65 K while only AHE hysteresis is seen at 80 K. (d) Magneto-optic Kerr effect (MOKE) measurements of 4 u.c. SRO film at 50, 60, 65, and 80 K. Only a simple rectangular hysteresis is observed in MOKE results.}
\label{fig:2}
\end{figure}

\section*{3. Interpretations for the hump-like behaviors in Hall measurement}
Figure 1a schematically illustrates Hall measurement data from 4 and 5 u.c. SRO thin films (without the ordinary Hall effect)\cite{sohn18}. Each Hall data consists of rectangular AHE curve and hump structures around the coercive fields. It is well-known that AHE is generally proportional to the magnetization\cite{nagaosa-review}, and the correlation between AHE and magnetization in SRO system has already been discussed\cite{fang03,mathieu04,khalifah07}. An interesting feature of the AHE in SRO is the thickness dependent sign of the anomalous Hall coefficient in the ultra-thin regime\cite{matsuno} as seen in Fig. 1a. Because of the sign change in AHE, detailed look of the Hall resistivity (that is, AHE plus hump structures) depends on the thickness. As seen in Fig. 1a, the Hall resistivity data for 4 and 5 u.c. look very different; while the hump structure appears to be similar, the AHE sign changes between the two cases.

The humps in the Hall data may be interpreted in two ways as mentioned above. The first interpretation, as illustrated in Fig. 1b, is based on the formation of the magnetic skyrmions. The Dzyaloshinskii-Moriya interaction from inversion symmetry breaking at the interface combined with the large SOC of Ru may lead to the formation of magnetic skyrmions which induce THE~\cite{sohn18}. Since magnetizations for 4 and 5 u.c. cases are similar, the skyrmion formation and thus the resulting THE effect remain similar between the two cases. Meanwhile, in spite of the similar magnetization, AHE changes the sign presumably due to location of the magnetic monopole relative to the Fermi energy~\cite{fang03}. The AHE and THE in Fig. 1b can result in the Hall effect shown in Fig. 1a.

On the other hand, the $\rho_{xy}$ in Fig. 1a may well be understood as a superposition of two (or more) AHE curves. Let us assume that there are two regions of 4 and 5 u.c. thick SRO which we may have due to the step-flow growth; a nominal 4 u.c. film may have dominantly 4 u.c. regions with some 5 u.c. regions, and the other way around for a nominal 5 u.c. film. The two regions have opposite AHE signs as mentioned above as well as different coercive field strengths as illustrated in Fig. 1c. The sum of the two curves in each case results in the $\rho_{xy}$ curves in Fig. 1a. Therefore, both interpretations successfully account for the $\rho_{xy}$ curves in Fig. 1a, which means that Hall measurement alone cannot resolve the issue and we need additional information to settle it down. Here, we note that the interpretation within the inhomogeneous AHE requires different coercive fields, which should be manifested in the magnetization v.s. magnetic field ({\it M-H}) curve as corresponding distinct steps or double hysteresis loops. Meanwhile, if the hump features are from magnetic skyrmions, we expect a simple single hysteresis loop in {\it M-H} curve since magnetic skyrmions have zero net magnetization. Therefore, magnetization measurements on SRO ultra-thin films should provide the answer to the issue.

\section*{4. magnetization measurement of SRO ultra-thin films}
In order to resolve the issue, we grew high quality 4 and 5 u.c. SRO films and carried out Hall effect measurements. Hall data taken at 10 K are plotted in Figs. 2a and 2b. Both of the data show clear AHE signal with opposite signs for the anomalous Hall coefficient. We thus only need to measure the magnetization as a function of the applied field. It has been shown that magnetization of ultra-thin films can be reliably obtained by MOKE measurements. Unfortunately, the coercive fields at 10 K ($\sim 1$ $T$) are larger than the highest field we can apply in the MOKE experiments ($\sim 0.1$ $T$), and the {\it M-H} curve could not be obtained over a wide enough field range. A way to circumvent the situation is to reduce the coercive field by raising the temperature - the coercive field diminishes as the temperature approaches the Curie temperature, $T_C$. Figure 2c shows $\rho_{xy}$ of 4 u.c. SRO thin film measured at 50, 60, 65 and 80 K. Note the much smaller coercive fields at these temperatures. In spite of much smaller coercive field values, the hump structures are clearly observed in the 50, 60, and 65 K data, whereas only small hysteresis is seen at 80 K which is close to $T_C$.

\begin{figure}[htbp]
\includegraphics[width=0.5\textwidth]{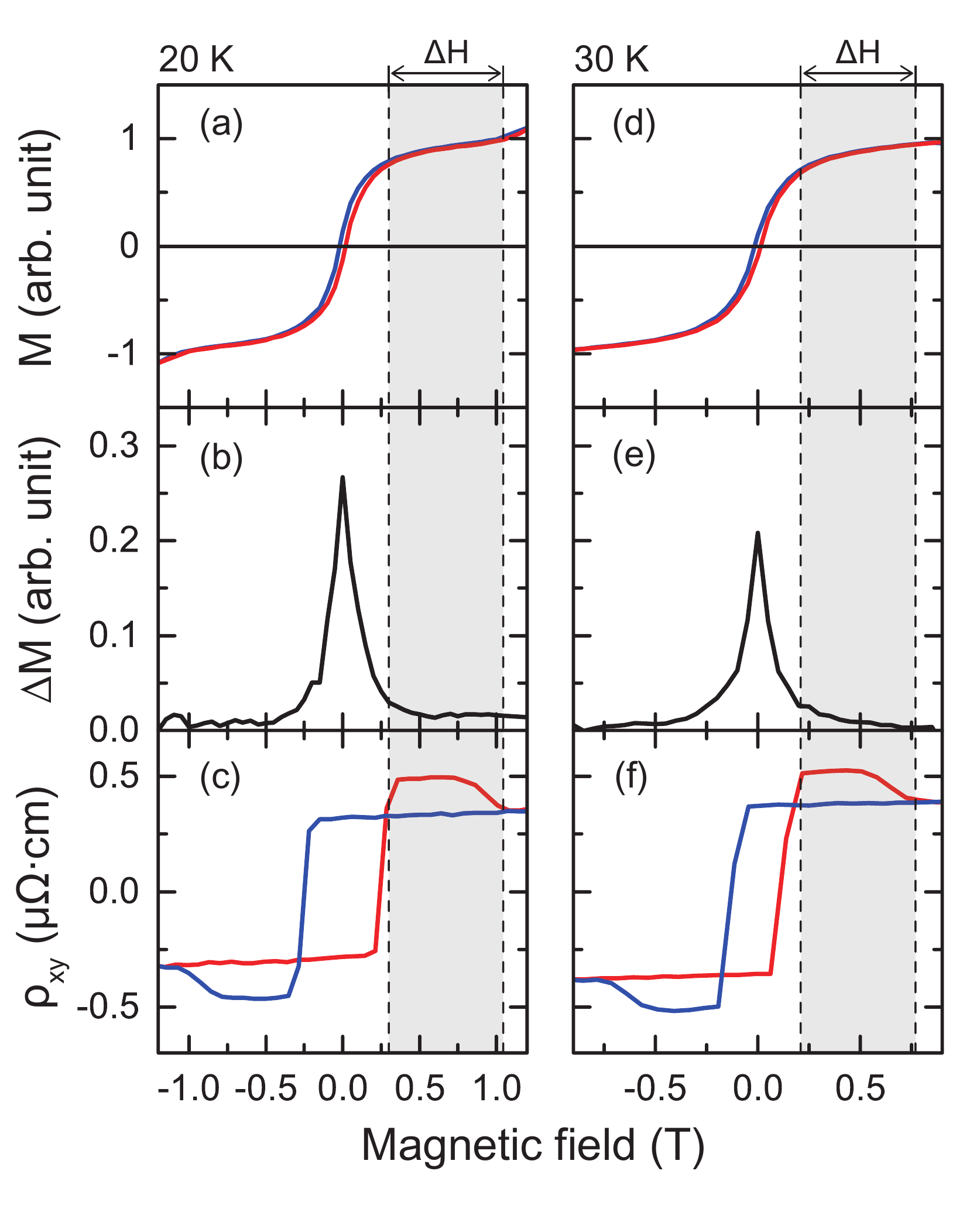}
\caption{(a) Normalized {\it M-H} curve of 4 u.c. SRO film measured with out-of-plane field direction at 20 K. Diamagnetic signal from the STO substrate has been subtracted. Red (blue) line represents the positive (negative) sweep direction of the applied magnetic field. Clear hysteresis is observed. (b) The difference between the two magnetizations measured with positive and negative sweep directions (difference between the two curves in (a)). (c) Hall resistivity $\rho_{xy}$ of the 4 u.c. SRO film measured at 20 K. (d), (e) and (f) Corresponding data set measured at 30 K. The dashed lines indicate the fields at which the magnetization nearly saturates ($\Delta M=0$) and THE disappears, respectively.}
\label{fig:3}
\end{figure}

Plotted in Fig. 2d are MOKE measurement results of 4 u.c. film at 50, 60, 65 and 80 K. One can clearly see hysteresis behavior in the Kerr rotation at 50, 60, and 65 K while it almost disappears at 80 K or near the $T_C$. While hump structure is clearly observed in the Hall data in Fig. 2c, the MOKE data in Fig. 2d show only single hysteresis loops without any sign of hump structures. The MOKE data therefore clearly show that the double AHE model cannot account for the hump structures observed in our 4 u.c. SRO films. On the other hand, the single hysteresis loop is not inconsistent with the view that the hump structure in the Hall data has a topological origin.

While the MOKE data already show that the double AHE model is inconsistent with the experimental data, studying the behavior of the magnetization at low temperatures is desired as the hump structure is much clear at low temperature. Magnetization measurement of ultra-thin film at low temperature in a field higher than the coercive field is possible with SQUID magnetometry. Although magnetization measurement by a SQUID has an issue of having the major contribution from the substrate than the film, it can still provide some useful information. Figures 3a and 3d show SQUID measurement results of 4 u.c. SRO film with an out-of-plane geometry at 20 and 30 K, respectively. Even though the magnetization data look different from what is expected from a hard magnet because of the contribution from the substrate, a hysteresis behavior due to the ferromagnetism of SRO thin film is discernible (See supplementary information for the raw data). The difference of the two magnetization curves (positive and negative sweeps), $\Delta M$, is plotted in Figs. 3b and 3e for 20 and 30 K, respectively. Meanwhile Figs. 3c and 3f present $\rho_{xy}$ of 4 u.c. SRO thin film taken at 20 and 30 K, respectively. Here, we pay attention to the shaded $\Delta H$ region between the vertical dotted lines. Although the hump feature is observed in the $\Delta H$ region, $\Delta M$ is nearly zero as $M$ is almost saturated. If the hump structure is due to inhomogeneous AHE stemming from inhomogeneous magnetization~\cite{rutgers18}, one has to see additional hysteresis in the $\Delta H$ region. Therefore, the magnetization data given in Fig. 3 are again inconsistent with the double AHE model.

At this point, it should be worthwhile mentioning a case in which both hump structures in the Hall data and a double hysteresis loop in {\it M-H} are observed. This is to show that our method is valid. In an earlier study, La$_{0.7}$Sr$_{0.3}$MnO$_3$/BaTiO$_3$/SrRuO$_3$ superlattice was prepared by PLD technique, and Hall effect measurement as well as magnetic characterization were performed on the superlattice system~\cite{Ziese18}. The results show hump features in the Hall data and a double hysteresis loop in the {\it M-H} curve. Since the superlattice system has two different ferromagnetic sources (La$_{0.7}$Sr$_{0.3}$MnO$_3$ and SrRuO$_3$), it is not surprising to see a double hysteresis loop in the {\it M-H} curve and resulting hump feature in the Hall effect data. Similar double hysteresis loop was also observed in SrRuO$_3$ (6 u.c.)/SrZrO$_3$/SrIrO$_3$/SrRuO$_3$ (18 u.c.) heterostructures in polar MOKE and MPMS measurements~\cite{Lena18}. In our ultra-thin SRO film case, however, we do not see any double hysteresis behavior in MOKE or SQUID results. We thus conclude that the double AHE model cannot account for the hump features in the Hall data from our ultra-thin SRO film.

\begin{figure}[htbp]
\begin{center}
\includegraphics[width=0.382\textwidth]{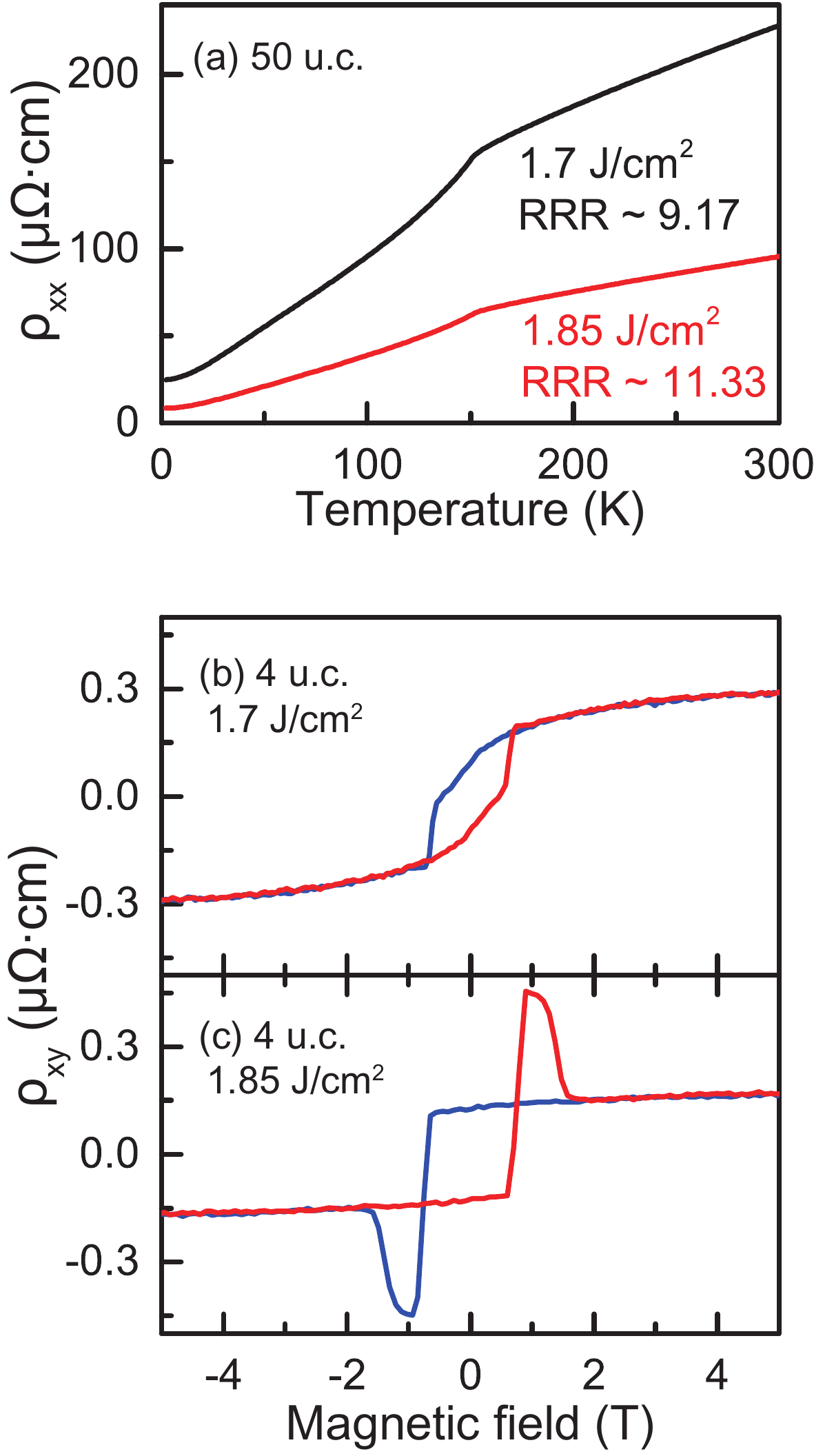}
\end{center}
\caption{(a) Temperature dependent resistivity for 50 u.c. SRO films grown under two different growth conditions. Black (red) line is from the film grown with a laser fluence of 1.7 (1.85) J/cm$^{2}$. (b) and (c) Hall resistivity of 4 u.c. SRO films grown under the two growth conditions. While no hump feature is observed for the 1.7 J/cm$^{2}$ film in (b), clear THE hump is seen in the data in (c) for 1.85 J/cm$^{2}$ film.}
\label{fig:4}
\end{figure}

Then, why do the results from different groups vary? In order to answer the question, we also grew SRO thin films under non-optimal growth conditions, intentionally lowering the film quality. Figure 4a shows temperature dependent resistivity data from two 50 u.c. SRO thin films grown with different laser fluence values of 1.7 and 1.85 J/cm$^{2}$ (here, we use 50 u.c. films for the characterization of the resistivity because these thick films do not have the low temperature up-turn). While the $T_C$ indicated by the kink in $R(T)$ is similar between the two cases, there is a clear difference in the resistivity value. The SRO film grown with 1.85 J/cm$^{2}$ has a higher RRR value of 11.33 compared to 9.17 for the film grown with 1.7 J/cm$^{2}$. In order to investigate the effect of the growth condition on the Hall effect, we grew 4 u.c. SRO thin films under the two growth conditions. Hall effect measurement results are presented in Figs. 4b and 4c. The hump features are clearly observed in the SRO thin film grown with 1.85 J/cm$^{2}$, whereas they are not seen for the film grown with the lower laser fluence. These results show that existence of the hump feature is very sensitive to the growth condition. The result of a previous study suggests that the RRR value is an indication of the Ru content in the system~\cite{siemons07}. Therefore, we may expect a higher Ru content in the $RRR=11.33$ sample than in the other sample. Based on these observations/results, we may suggest that the difference between Hall measurement results from different groups may be from the variation in the Ru content which originates from different growth conditions. While the exact relation between the hump feature and Ru content should be left for future studies, it is clear that the hump feature appears only for films with high RRR values.

\section*{5. conclusion}
In conclusion, we performed Hall effect and magnetization measurements on SRO ultra-thin films grown on STO $(001)$ substrate. The Hall effect data show AHE as well as clear hump features. MOKE and SQUID measurements on 4 u.c. SRO film did not show a double step hysteresis loop but a simple single step loop. This is a clear evidence against the view that the hump structures are due to regions with different magnetic properties (that is, AHE). Therefore, we conclude that the hump-like behavior in our SRO thin films cannot be explained within the inhomogeneous AHE model and must have topological origin. In addition, we show that the hump behavior in the Hall resistivity can vary depending on the growth condition, which we attribute to the growth condition dependent Ru content in the film.

\acknowledgments

This work is supported by IBS-R009-G2 through the IBS Center for Correlated Electron Systems. S.H.C. was supported by Basic Science Research Program through NRF (2016K1A3A7A09005337). J.W.C. acknowledges KIST Institutional Program.


\end{document}